%% file: main.tex
\documentclass[a4paper,11pt]{article}

\usepackage{jinstpub} 

\usepackage{booktabs}
\usepackage{multirow}
\usepackage{lineno}

\title{\boldmath Pulse shape discrimination using a convolutional neural network for organic liquid scintillator signals}

\input{authors}

\abstract{A convolutional neural network (CNN) architecture is developed to improve the pulse shape discrimination (PSD) power of the gadolinium-loaded organic liquid scintillation detector to reduce the fast neutron background in the inverse beta decay candidate events of the NEOS-II data.
A power spectrum of an event is constructed using a fast Fourier transform of the time domain raw waveforms and put into CNN. 
An early data set is evaluated by CNN after it is trained using low energy $\beta$ and $\alpha$ events.
The signal-to-background ratio averaged over 1-10 MeV visible energy range is enhanced by more than 20\% in the result of the CNN method compared to that of an existing conventional PSD method, and the improvement is even higher in the low energy region.
}

\arxivnumber{2211.07892} 

\collaboration[c]{(NEOS-II collaboration)}

\begin{document}
\maketitle
\flushbottom

\input{1.intro}
\input{2.data_preparation}
\input{3.cnn_train}
\input{4.results}

\input{5.summary}

\acknowledgments
This work is supported by the National Research Foundation of Korea (NRF-2022R1A2C1009686). NEOS-II is supported by Institute for Basic Science (IBS-R016-D1 and IBS-R016-D1-2018-B01). Computational works for the research were performed on the data analysis hub, Olaf in the IBS Research Solution Center.

\bibliography{references}

\end{document}

%% file: authors.tex
\author[a]{K. Y. Jung}
\author[b]{B. Y. Han}
\author[c]{E. J. Jeon}
\author[a]{Y. Jeong}
\author[d]{H. S. Jo}
\author[b]{J. Y. Kim}
\author[f]{J. G. Kim}
\author[c,e,g]{Y. D. Kim}
\author[c]{Y. J. Ko}
\author[c,g]{M. H. Lee}
\author[c]{J. Lee}
\author[d]{C. S. Moon}
\author[c,1]{Y. M. Oh\note{corresponding author}}
\author[h]{H. K. Park}
\author[c]{S. H. Seo}
\author[a]{D. W. Seol}
\author[a]{K. Siyeon}
\author[b]{G. M. Sun}
\author[i]{Y. S. Yoon}
\author[f]{I. Yu}

\affiliation[a]{Department of Physics, Chung-Ang University, Seoul, 06974, Korea}
\affiliation[b]{HANARO Utilization Division, Korea Atomic Energy Research Institute, Deajeon, 34057, Korea}
\affiliation[c]{Center for Underground Physics, Institute for Basic Science (IBS), Daejeon, 34126, Korea}
\affiliation[d]{Department of Physics, Kyungpook National University, Daegu, 41566, Korea}
\affiliation[e]{Department of Physics and Astronomy, Sejong University, Seoul, 05006, Korea}
\affiliation[f]{Department of Physics, SungKyunKwan University, Suwon, 16419, Korea}
\affiliation[g]{IBS School, University of Science and Technology (UST), Daejeon, 34113, Korea}
\affiliation[h]{Department of Accelerator Science, Korea University, Sejong, 30019, Korea}
\affiliation[i]{Ionizing radiation metrology group, Korea Research Institute of Standards and Science, Daejeon, 34113, Korea}

\emailAdd{yoomin@ibs.re.kr}

%% file: 1.intro.tex
\section{Introduction}
\label{sec:intro}

The signature of a prompt positron and a delayed neutron capture signal pair from the inverse beta decay~(IBD) reaction~\cite{Vogel:1999zy} enables us to detect the electron antineutrino's weak interaction signal buried in a large number of background events. One of the major backgrounds which mimic the IBD signal pair in the organic scintillation detector is the fast neutron's moderation and capture. In the case of the reactor experiments, it is preferred to deploy the detector underground to reduce the cosmogenic neutron rate. However, very short baseline experiments to solve the so-called Reactor Antineutrino Anomaly~(RAA)~\cite{Mention_2011} must deploy their detectors at distances within tens of meters from the reactor cores, where only a minimal shield from the cosmogenic background is provided. Although most cosmogenic fast neutron background, which is not related to the reactor operation, can be eliminated by subtracting the reactor-off background data, there may remain uncertainty from the variation of the fast neutron flux depending on the atmospheric conditions~\cite{Davis:2016gkz,STEREO:2019ztb}. To reduce the systematic uncertainty that embeds this time-varying background  improve statistical precision, it is better to reduce the fast neutron background as much as possible. With the help of digital signal processing, the fast-$n$ background can be mitigated using the pulse shape discrimination~(PSD) at the analysis level. PSD is based on the fact that the scintillation pulse shape of the fast-$n$'s moderation, mainly by proton recoil, is different from that of $\gamma$'s or $e^{+/-}$'s electron recoil signal. Since the difference in the pulse shapes lies primarily in the slow components of the scintillation, the ratio of the pulse tail area to the total area ($F_{\textrm{tail}}$) in the time domain is generally used as the PSD parameter. The discrimination power worsens at the lower visible energy side due to poorer photostatistics and signal-to-noise ratio. Other methods using the Fourier transform have been tested in order to get over the degradation by high-frequency noises~\cite{5485131,BALMER2015146,HUBBARD201964}.

While the traditional PSD techniques mainly focus on finding principle components and constructing simple parameters, convolutional neural networks (CNN) can take the raw signals as the input data. For decades, CNN architectures have been introduced and developed for image classification~\cite{Lenet5,NIPS2012_c399862d,https://doi.org/10.48550/arxiv.1409.1556,https://doi.org/10.48550/arxiv.1409.4842,https://doi.org/10.48550/arxiv.1512.03385}. In the neutrino physics field, CNN has been studied for the classification of neutrino interaction types, particle identification, background rejection, and so on~\cite{Aurisano:2016jvx,DUNE:2020gpm,MicroBooNE:2016dpb,MicroBooNE:2018kka,MicroBooNE:2020yze,NEXT:2016ire,DBLP:journals/corr/abs-1809-06166,7838264,Domine:2019zhm,KamLAND-Zen:2016pfg,Griffiths:2018zde,MINERvA:2018smv}.

NEOS~\cite{NEOS:2016wee} used a linear-alkylbenzene (LAB) based gadolinium (Gd)-loaded liquid scintillator, into which a 10\% volume of a di-isopropylnaphthalene (DIN) based liquid scintillator was mixed to enhance the PSD capability~\cite{Kim_2015,NEOS:2015dzs}. As a result of the phase-I measurement using the $F_{\textrm{tail}}$ method, the background was reduced by 70\%, and a signal-to-background ratio of 22 was achieved in the 1-10 MeV prompt energy range~\cite{yoomin:2018nu, NEOS:2016wee}.  We have reported a possible PSD improvement using a CNN applied to a test data set consisting of fast-$n$ and $\beta/\gamma$ background events~\cite{NEOSII:2020gai}. The time domain waveforms used to calculate $F_{\textrm{tail}}$ were given as input data for the CNN training and evaluation.

We have continued the study for further PSD improvement and to see the impact on the experimental result of NEOS-II, which is presented in the following sections. 
First, data processing for the conventional PSD method utilizing the tail charge and the construction of the frequency domain power spectrum using fast Fourier transform (FFT) in the NEOS-II analysis are described in detail in section~\ref{sec:data}. In section~\ref{sec:cnn} the CNN method using the FFT power spectrum is explained. We will apply both the conventional and CNN methods on the NEOS-II IBD candidate dataset and compare the results in section~\ref{sec:result}.

%% file: 2.data_preparation.tex
\section{Data preparation for PSD}
\label{sec:data}

\subsection{A conventional method}
\label{subsec:psd_ftail}

\begin{figure}[!th]
\centering
\fbox{\includegraphics[width=0.95\textwidth]{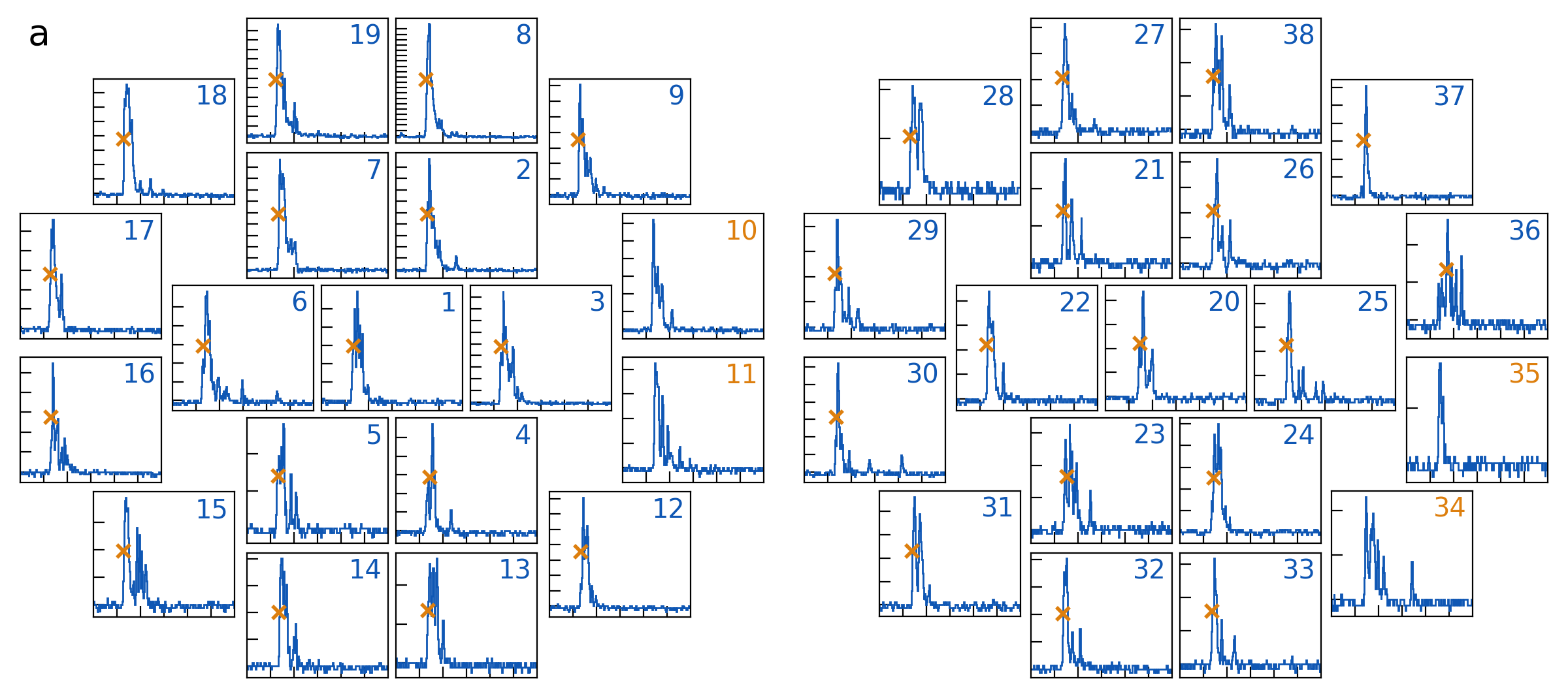}}\\\vspace{10pt}
\includegraphics[width=0.44\textwidth]{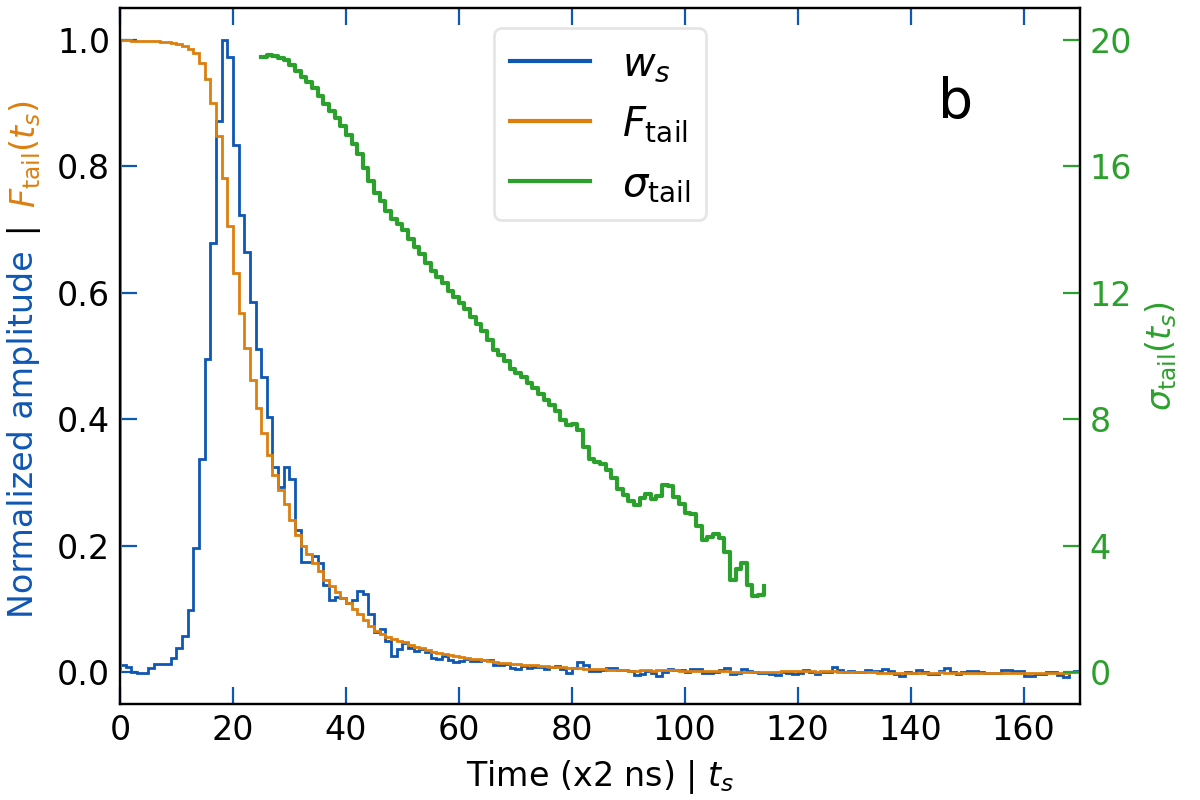}
\hspace{0.04\textwidth}
\includegraphics[width=0.44\textwidth]{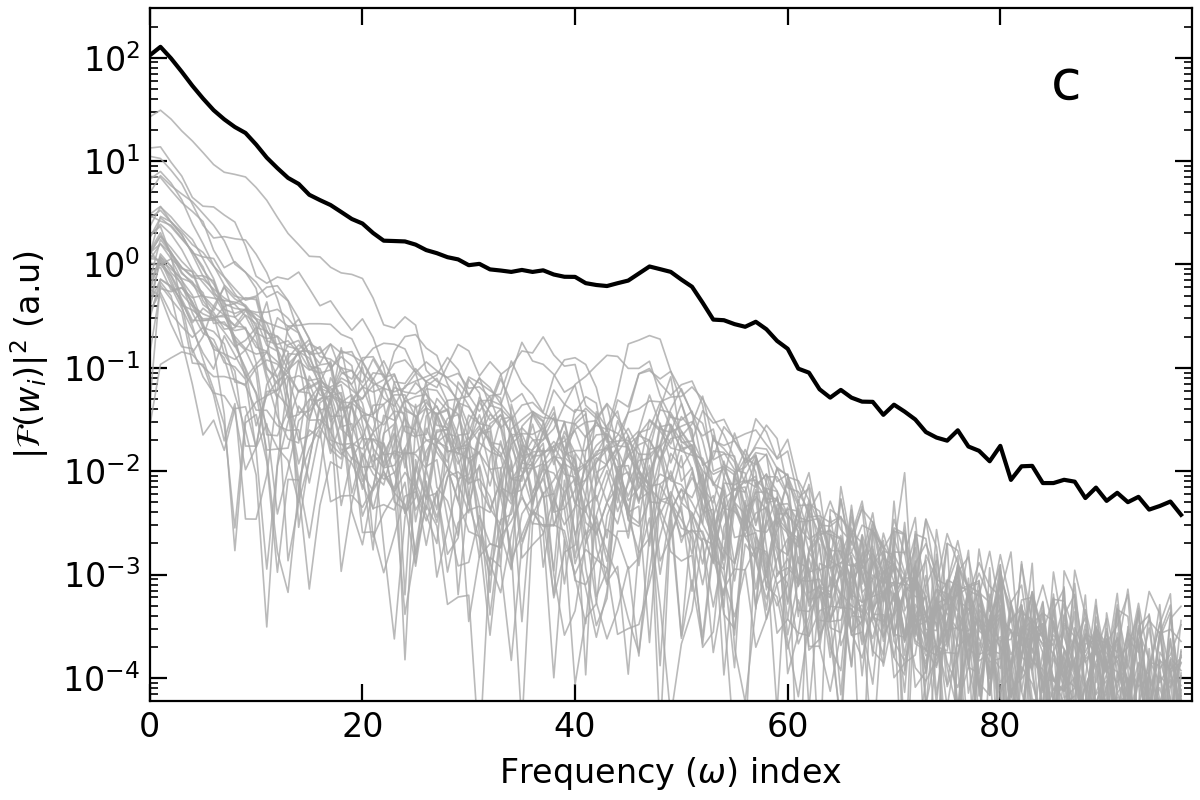}
\caption{An example of the raw data processing of an event for PSD. (a) Raw waveforms ($w_{i}$'s) from 38 PMTs. All $x$-axes share the same time range (0-480 ns), and a division in $y$-axes stands for 5 mV. The pulse time ($t_{p}$), drawn as an orange cross, is defined at the half-maximum point in the rising part. (b) Waveforms whose $t_{p}$'s are properly found are synchronized and accumulated to produce a synchronized waveform (blue). $F_\mathrm{tail}$~(orange) and $\sigma_{\mathrm{tail}}$ (green) are then calculated for each tail-starting time bin. (c) Power spectra of 38 PMT waveforms (thin gray) and their sum (thick black). See the text for a detailed description.}
\label{fig:rawwaves}
\end{figure}

A signal in the liquid scintillator is detected by 2$\times$19 photomultiplier tubes (PMTs) in the NEOS detector~\cite{NEOS:2016wee}. Each PMT signal is digitized via a flash analog-to-digital converter (FADC) module which has a 12-bit resolution for 2.5 volts peak-to-peak dynamic range and a 500 MHz sampling rate. The digitized 38 waveforms in a 480 nanosecond window are then stored as the raw data of an event. An example of the raw waveforms of an event is shown in figure~\ref{fig:rawwaves}a.

\begin{figure}[!htb]
    \centering
    \includegraphics[width=0.98\textwidth]{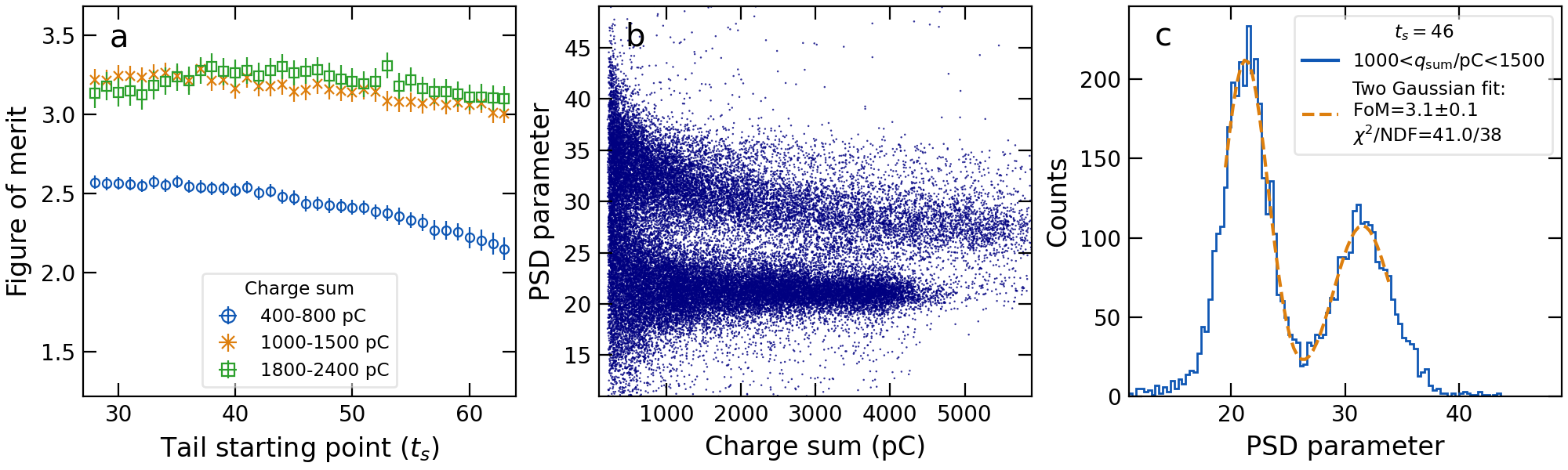}
    \caption{(a) Discrimination powers (FoM) at different tail starting points for different charge sum ranges; (b) PSD parameter, $\sigma(t_{s}=46)$ versus charge sum; (c) $\sigma(t_{s}=46)$ distribution and its fit to a two-Gaussian function for events in 2-3 MeV range (charge sum of 1000-1500 pC range). Data shown here are the prompt events, followed by $n$-capture-on-Gd events in the reactor-off period data.}
    \label{fig:opt_ftail}
\end{figure}

For the pulse shape analysis in the time domain, a synchronized waveform ($w_{s}$) is constructed (see figure~\ref{fig:rawwaves}b). Considering the detector size and the PMT transit time spread, signal timings in channels can be different by several nanoseconds. First, we look at the rising part of the $i$th PMT's raw waveform between the maximum point and the last point exceeding the pedestal root mean square before the maximum. When the rising part consists of more than 3 points, we define a pulse time ($t_{p}^{i}$) of the waveform at the half maximum point found by fitting $\left(t,\,\left[-\log\left(w_{i}(t)/w_{i}^{\mathrm{max}}\right)\right]^{1/2}\right)$ to a linear function, where $w_{i}(t)$ stands for the waveform amplitude at a time bin $t$ for $i$th PMT, and $w_{i}^{\mathrm{max}}$ for its maximum amplitude. While the total charge of an event ($q_{\mathrm{sum}}$), a measure of the energy deposit in the detector~\cite{yoomin:2018nu}, is calculated from the sum of the integrated area of all 38 raw waveforms, a synchronized waveform ($w_{s}$) of the event is formed by accumulating the waveforms of which $t_{p}$'s are defined:
\begin{equation}
\label{eq:wsync}
w_{s}(t) = \sum_{i}w_{i}\left(t_{p}^{i}-25+t\right),
\end{equation}
where $t$ is the time bin index from 1 to 170. The selection of 170-time bins allows it to contain most of the main signal information and exclude the pedestal and possible afterpulses. The tail-to-total charge ratio, $F_{\mathrm{tail}}$, is calculated from $w_s$. In the NEOS-II analysis, we have developed another PSD parameter, which is defined as the spread of mean-time in the tail range of $w_{s}(t)$ as follows:
\begin{equation}
    \label{eq:psd_param}
    \sigma_\mathrm{tail}(t_s) = \sqrt{\sum_{t=t_{s}}^{170} \frac{t^{2}\cdot w_{s}(t)}{A_{\mathrm{total}}}-\sum_{t=t_{s}}^{170}\left[ \frac{t\cdot w_{s}(t)}{A_{\mathrm{total}}}\right]^{2}},
\end{equation}
where $t_{s}$ is the tail starting bin, and $A_{\mathrm{total}}=\sum_{t=1}^{170}w_{s}(t)$ is the total area of the synchronized waveform.
The conventional optimization of PSD is finding $t_{s}$ which maximizes the discrimination power, i.e., the figure of merit (FoM), between $\beta/\gamma$ and fast-neutron/$\alpha$ events:
\begin{equation}
    \label{eq:fom}
    \mathrm{FoM} = \frac{\left|m_{\gamma} - m_{n}\right|}{\sqrt{\sigma_{\gamma}^{2}+\sigma_{n}^{2}}},
\end{equation}
where $m$ and $\sigma$ denote the Gaussian mean and width of a type of scintillation event--$\gamma$ for $\beta/\gamma$ and $n$ for fast-neutron/$\alpha$, respectively. Figure~\ref{fig:opt_ftail}a shows how FoM and optimum $t_{s}$ values change for different $q_{\mathrm{sum}}$ ranges.

It is natural to have a small FoM in the low $q_{\mathrm{sum}}$ range, as shown in figure~\ref{fig:opt_ftail}, because of a small number of photons, which may be further degraded by choice of waveforms with valid $t_{p}$'s. An example of a low energy ($\sim$1 MeV, or charge~500 pC) event in figure~\ref{fig:rawwaves} shows that  $t_{p}$ cannot be found for some channels (10, 11, 34, and 35). 

\subsection{Utilization of the fast Fourier transform}
\label{ssec:fft}
\begin{figure}[!htb]
    \centering
    \includegraphics[width=0.96\textwidth]{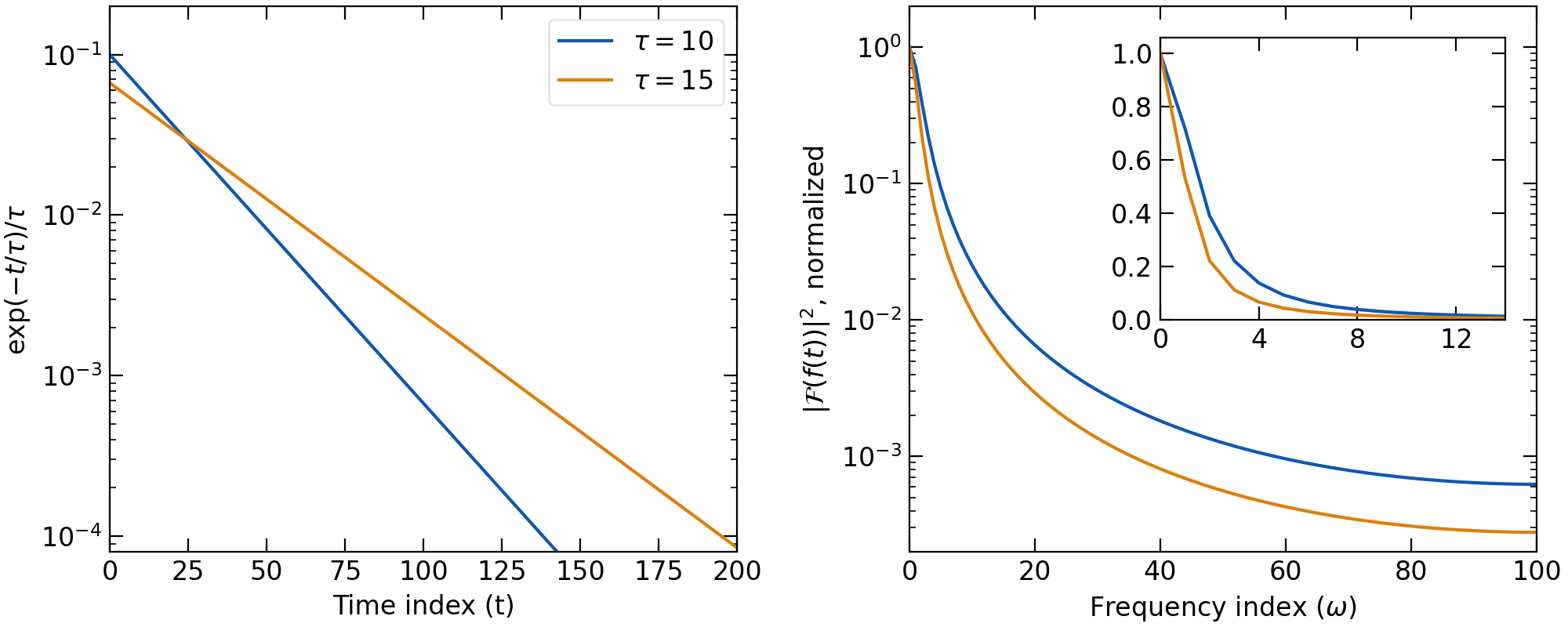}
    \caption{Two exponential functions with different decay times (left) and their normalized squared amplitudes in the frequency domain after a fast Fourier transform (right). The low frequency region is shown on a linear scale in the inset.}
    \label{fig:fft_exp}
\end{figure}

Generally, a proton recoil event shows a larger decay time in the slow decaying component in the organic scintillator than an electron recoil event.
It means that, when the total sizes of the two pulses are the same, the pulse of the former varies smaller in the decaying part and will show a smaller amplitude at the corresponding frequency range after the Fourier transform.
A simple example is shown in figure~\ref{fig:fft_exp}.
A normalized power spectrum in the frequency domain, denoted as $\mathcal{P}(\omega)$, can be constructed by adding the amplitude-squared of the Fourier transform ($\mathcal{F}$) of each PMT's waveform:
\begin{equation}
\label{eq:fft}
\mathcal{P}(\omega)=\sum_{i=1}^{38} \left|\mathcal{F}\left(w_{i}(t)\right)\right|^{2}.
\end{equation}
The fast Fourier transform is used since the waveforms have a finite length and are discretized in time. The waveforms in the time bins between the 25th from the start and the 20th from the end are used for the transform, to exclude useless pedestal parts and to avoid possible afterpulses, respectively. Figure~\ref{fig:rawwaves}c shows each of the 38 power spectra ($|\mathcal{F}(w_{i}(t))|^{2}$) and their sum ($\mathcal{P}(\omega)$). The zeroth term of $\mathcal{P}(\omega)$, $\mathcal{P}(0)$, is proportional to the quadratic sum of the integrated area of the time domain signal and can be used as the denominator for normalization. One advantage of using the frequency domain power spectrum is that one doesn't have to worry about time information, i.e., finding pulse times that may introduce an additional error so that all 38 PMT signals can be used. Another advantage is data reduction. Neglecting the phase terms leaves a half of the number of original time bins. It helps save computing power when the spectrum is used as an input for CNN.

\subsection{Preparation of data samples}
\label{ssec:data_samples}

There are many ways to get pure electron recoil events in the NEOS data, while it is almost impossible to sort out pure proton recoil events in the wide energy range of interest ($\sim$1$<E$ /MeV$<$10).
Gamma-event dominant data can be found in most of the source calibration data. Neutron-capture $\gamma$ events depositing energies up to around 8 MeV in the detector can be found in the delayed coincident event(s) of $^{252}$Cf or PoBe source calibration data. In the case of $^{252}$Cf, multiple neutrons' moderation and capture may happen in an event time window, so that pile-up signals can pollute some waveforms. In the PoBe case, the source emits a fast neutron with or without a 4.4 MeV $\gamma$. A proton recoil by the fast neutron's moderation leaves a visible energy of less than 2 MeV, and sometimes a pile-up happens by the coincidence of the 4.4 MeV $\gamma$ and the moderation of the neutron. Another drawback is that events mostly occur at around the top and middle of the horizontal cylindrical detector due to the source position, so the scintillation photons are observed not uniformly or symmetrically for 38 PMTs but mainly by PMTs at the top part. Other calibration sources, $^{22}$Na, $^{60}$Co, and $^{137}$Cs, provide symmetric charge distributions of $\gamma$ events for 38 PMTs but are limited to the low energy side.

In the background data, events in the muon veto time window, 150 microseconds after a muon counter event, have rich information, although they are unselected in the IBD reconstruction. First, the Michel electron~\cite{Michel:1949qe} can be found by a delayed coincidence with stopping muon events. Some of them may coincide with afterpulses of the large muon signals; the contribution of the afterpulse becomes negligible at high energy. Therefore the Michel electron event above 4 MeV can be used as an electron recoil sample. Second, there are fast neutrons following a muon. We can find the moderation of fast neutrons leaving proton recoil signals and the electron recoil signals of $\gamma$ events by the neutron captures (figure~\ref{fig:opt_ftail}). In this case, both types of recoil events leave visible energy wide enough for our interest, but the events cannot be clearly classified into the recoil types without PSD.

Consecutive $\beta-\alpha$ decays of $^{214}$Bi and $^{214}$Po, originating from the radon contamination in the liquid scintillator, provide clean samples of electron and proton recoil events, respectively, at the low visible energy range. Thanks to the short half-life of $^{214}$Po (164 microseconds) and the full energy deposit from its $\alpha$ decay, the pair of decay events can be selected by a 300 microsecond coincidence time window, a cut on the PMT charge sum around 500~pC for the delayed event, and a loose $F_\mathrm{tail}$ selection, as shown in figure~\ref{fig:BiPo}a. Then a fine selection of the charge and energy values is made for the $\alpha$ events.

\begin{table}[!htb]
\begin{minipage}{\textwidth}
\centering
\caption{Usable data samples for the PSD analysis. Calibration sources have typical radioactivities of $\mathcal{O}(10^{2})$~Bq.}
\label{tab:data_samples}
\begin{tabular}{@{ }ccccl@{ }}
\toprule
   Data type  & Source & Recoil & Visible energy (MeV) & Comment\\
\midrule
   \multirow{6}{*}{Calibration}&  $^{22}$Na,$^{60}$Co,$^{137}$Cs & $e$ & $<\sim$2.5 & Low energy \\
   \cmidrule{2-5}
   & \multirow{2}{*}{$^{252}$Cf} & $e$ & $<$10 &
    \multirow{2}{*}{Pile-up of multiple neutrons} \\
    \cmidrule{3-4}
   &  & $p$ & DR$^{*}$ \\
   \cmidrule{2-5}
   & \multirow{2}{*}{PoBe} & $e$ & 1-10 & Asymmetric source position\\
   \cmidrule{3-4}
   & & $p$ & <2 &  Pile-up of prompt fast-$n+\gamma$\\
\midrule
   \multirow{6}{*}{Background} & Michel electron & $e$ & $>$4 & High energy, $\sim$0.7 Hz \\
   \cmidrule{2-5}
    & \multirow{2}{*}{$\mu$-induced $n$} & $e$ & 1-10 & Similar to IBD\\
    \cmidrule{3-4}
    & & $p$ & DR$^{*}$ & $\mathcal{O}(10^{2})$~events/day\\
    \cmidrule{2-5}
    & \multirow{2}{*}{Bi($\beta$)-Po($\alpha$)} & $e$ & $<$3 & Low energy, clean samples\\
    \cmidrule{3-4}
    &  & $p$ & $\sim$1 & $\mathcal{O}(10^{4})$~events/day\\
\bottomrule
\end{tabular}
{\footnotesize ${}$~$^{*}$DR: in the whole dynamic range of the detector: 0.6-20 MeV.$\;$}\hspace*{\fill}
\end{minipage}
\end{table}

%% file: 3.cnn_train.tex
\section{PSD using Convolutional Neural Network}
\label{sec:cnn}

\subsection{General framework}
\label{subsec:cnnframework}
\begin{figure}[htb!]
\includegraphics[width=1\textwidth]{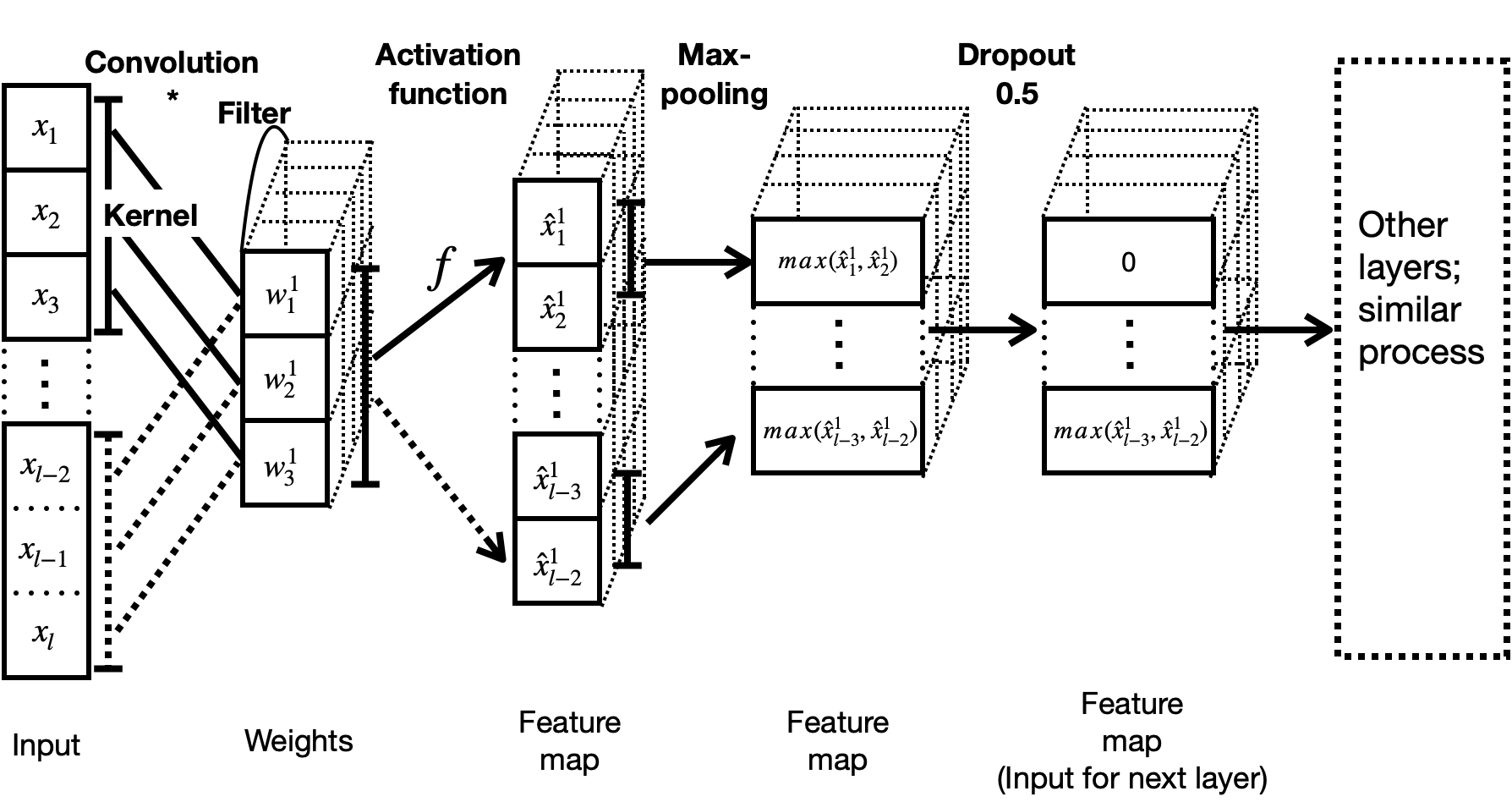}
\caption{An example of CNN framework and hyperparameters' role. This cartoon shows three kernels, four filters, max pooling of two, and a dropout of 0.5. The index is the same as in equation~\ref{featuremap}}
\label{fig:CNNframe}
\end{figure}

A neural network optimizes weight parameters that transform input data within the hidden layers of the network.
The traditional deep learning connects the weight parameters to the full input shape, which may cause overfitting and a lack of generalization.
The weight parameters in CNN are connected to a subset of the input shape and shared by the whole input to avoid overfitting and to improve generalization~\cite{Goodfellow-et-al-2016}.
There is a series of hyperparameters when CNN optimizes the weight parameters, for example, kernel, filter, stride, activation function, pooling, dropout, etc. With these hyperparameters, CNN extracts the $j$'th feature map, $\hat{x}^{l}_{j}$, defined as 
\begin{equation}\label{featuremap}
\hat{x}^{l}_{j} \equiv f\left(\overset{m}{\underset{i=1}{\sum}}x_{i+j}w^{l}_{i}+b^{l}_{j}\right),
\end{equation}
where $m$ is a given kernel size, $l$ is the number of filters, $x_{i}$ is the $i$'th input, $f$ is an activation function, b is a bias, and $w$ represents weights. The kernel is the size of shared parameters in a layer. The weights in the kernel are combined with input data in an activation function to extract the feature map. A filter is a number of kernels and a depth of feature map to learn various features. The stride indicates the amount of movement for the kernel over the input. The activation function is an operator on the linear combination of input and weight parameters to produce features; ReLU (rectified linear unit), hyperbolic tangent, and sigmoid functions are used in this study. 
Pooling and dropout are used to prevent overfitting which is specialized for CNN in specific samples. Pooling is a way to reduce the size of a feature map, such as max pooling, average pooling, and stochastic pooling. In this study, max pooling is adopted, choosing a max value of the neighbor in the feature map. Dropout removes a part of the feature map to prevent overfitting and improve the generality of CNN. Figure~\ref{fig:CNNframe} shows the general framework of CNN and each hyperparameter's role. When the CNN has more than one layer, the feature map after the convolutional layer becomes the input of the next layer. 

For the supervised binary classification, each input getting into the CNN training needs to be labeled as the true information. In our case, the label for an electron recoil ($\beta/\gamma$) event is assigned as 1, and 0 for a proton recoil ($\alpha /$fast-$n$) event. Each event is evaluated with a score of a CNN result between 0 and 1, indicating how close the event is to one label or the other. A loss is calculated from the difference between the score and the label. The binary entropy function is defined as: 
\begin{equation}
\label{eq:lossfunction}
L(w)=-\frac{1}{N}\sum_{j=1}^{N} \left[y_{j}\cdot log(\hat{y}_{j})+(1-y_{j})\cdot log(1-\hat{y}_{j})\right],
\end{equation}
where $\hat{y}_{j}$ and $y_{j}$ are the scores as a function of weights and the label for the $j$-th input. This loss function can diverge infinitely when the difference between the label and score gets larger. CNN updates the optimized weights in the direction of gradient descent of the loss function by backpropagation.

\subsection{CNN training and architecture}
\label{subsec:CNNarchitecturetraining}

\begin{figure}[htb!]
\centering
\includegraphics[width=1\linewidth]{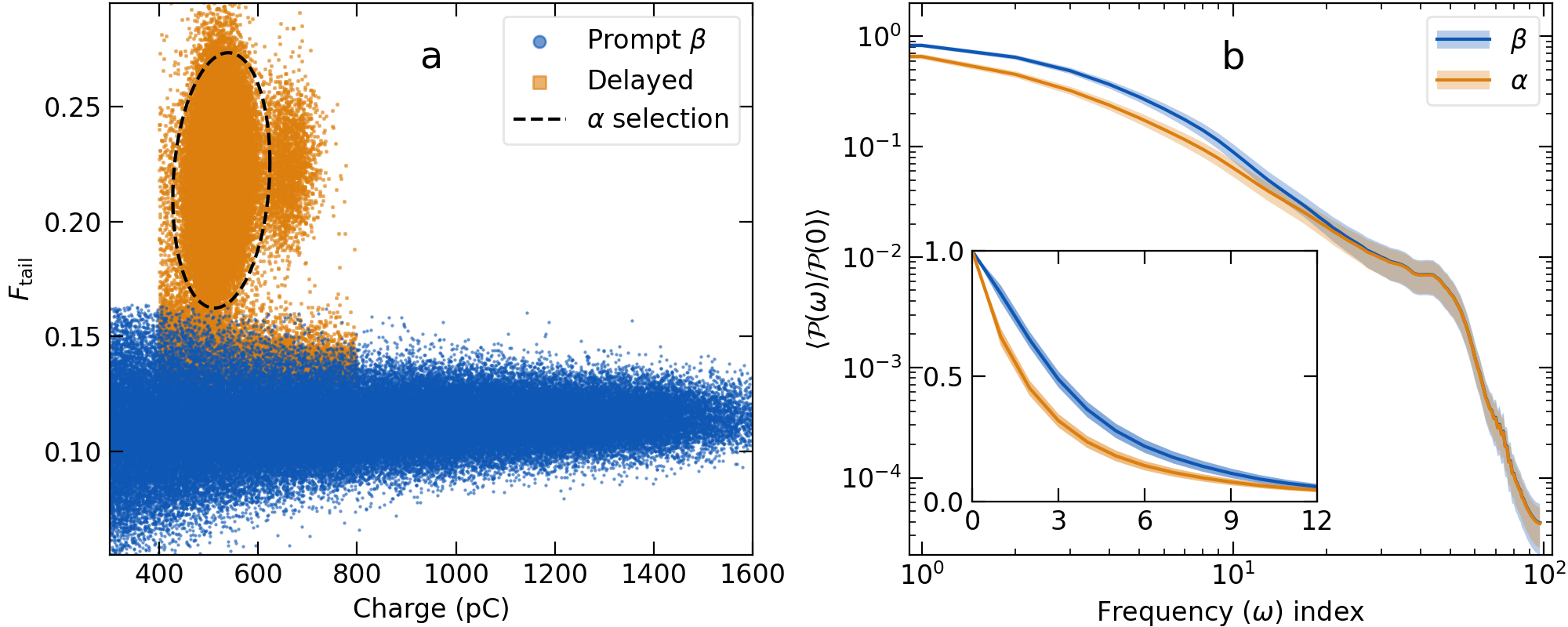}
\caption{Consecutive $^{214}$Bi's $\beta$- and $^{214}$Po's $\alpha$- decay event candidates. (a) $F_\mathrm{tail}$ versus charge distribution. The $\beta$ events are selected for $F_{\mathrm{tail}}<0.16$, and the $\alpha$ events are selected inside the black dashed ellipse. (b) Averages (solid curves) and the standard deviations of normalized power spectra for $\beta$ (blue) and $\alpha$ (orange) events where $\beta$ events are selected to have a similar mean charge with the $\alpha$ events.}
\label{fig:BiPo}
\end{figure}

It is our best choice to use the Bi-Po events described in section~\ref{ssec:data_samples} as the training input for CNN because they provide us with the cleanest samples of the electron and proton recoil events at low energy, where an efficient rejection of the fast-$n$ background in the IBD candidate events is difficult by the conventional PSD method. Similarities of the synchronized waveforms between $\beta$ and $\gamma$ and between $\alpha$ and fast-$n$ are already shown in our previous work~\cite{NEOSII:2020gai}. The normalized power spectrum, $\mathcal{P}(\omega)/\mathcal{P}(0)$, which has 98 data points with its label, is put into the CNN input layer. The reasons why the FFT spectrum is used instead of the time-synchronized waveform are; (1) to fully utilize the 38 PMT waveforms as explained in section~\ref{ssec:fft}, (2) to avoid the possible bias caused by setting the pulse time at a given time point for the synchronized waveform, and (3) to save computing resources by reducing the input data shape from 170 of the time synchronized waveform to 98 of the FFT power spectrum. Figure~\ref{fig:BiPo}a shows the selected $\beta$ and $\alpha$ events in the $F_{\mathrm{tail}}$ versus charge plane, and figure~\ref{fig:BiPo}b shows the averages of their normalized power spectra. One can see that the main difference between the two power spectra lies in the low-frequency range.

Considering separation power and generality, we tried constructing CNN architectures as simply as possible using large strides and a few filters~\cite{Asurvey}. Because there is no mathematical formula to find an optimized CNN architecture, the appropriate architecture and hyperparameters are found by trial and error~\cite{rala2021neural}.
Out of many different CNN architectures that are tried, the ones with relatively simple layers and structures for 1D CNN show satisfactory performances in signal classification compared to more complicated architectures in terms of accuracy and loss. One of the simplest architectures that offers the lowest loss and the best accuracy for the evaluation data samples is chosen.

\begin{figure}[!t]
\centering
\includegraphics[width=0.95\textwidth]{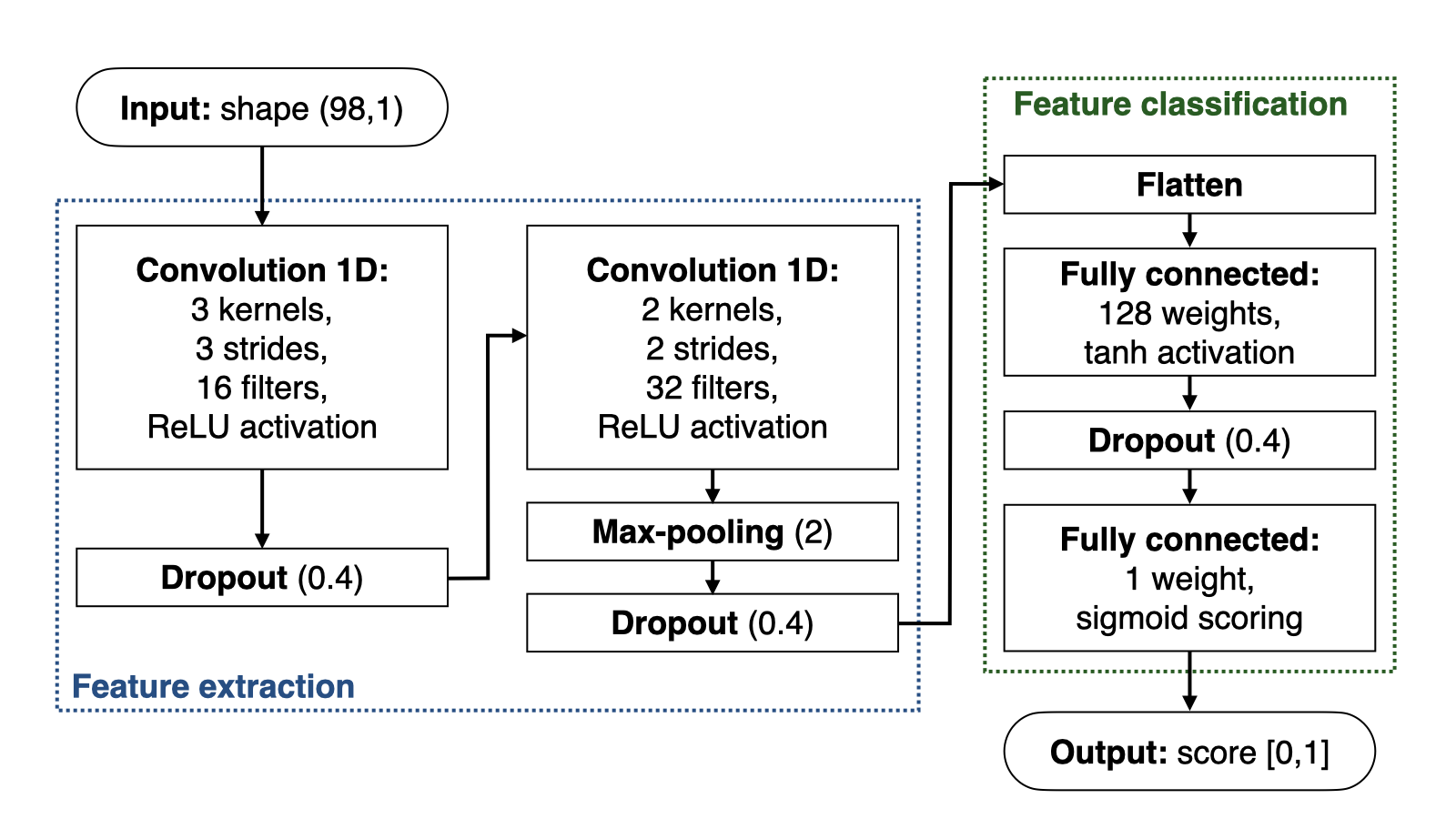}
\caption{The CNN architecture design for this work. See the text for the detailed description.
}
\label{fig:CNNarchitecture}
\end{figure}

The CNN architecture of this work is drawn in figure~\ref{fig:CNNarchitecture}. Keras~\cite{chollet2015keras} and Tensorflow~\cite{DBLP:journals/corr/AbadiBCCDDDGIIK16} are used for CNN training on a computer with an NVIDIA GeForce RTX 2080 GPU. Data for training are divided into 80\% for the training set and 20\% for the validation set. First, every batch consisting of 128 events, each of which has 98 data points, in the training data set goes through the first convolutional layer consisting of 16 kernels of size three and stride 3. Its outputs are then put into the second convolutional layer with 32 kernels of size two and stride 2. During these feature extraction processes, the ReLU activation function is applied to every convolutional layer. A max-pooling and a dropout are applied to the second layer and every convolutional layer, respectively, to prevent overfitting. The feature maps extracted through the convolutional and pooling layers are then fully connected to 128 weights by a $\tanh$ activation function for the feature classification.
Finally, the result from the fully-connected layer turns into the score by the sigmoid function.
The weights are updated in the direction of loss minimization for the next batch, using the binary cross entropy (equation~\ref{eq:lossfunction}) and an Adam optimizer~\cite{https://doi.org/10.48550/arxiv.1412.6980}. An epoch is completed when the entire batches pass through the CNN training, where computing time for each epoch takes less than 30 sec.

\subsection{Training result}
\label{subsec:Trainingresult}
\begin{figure}[htb!]%
\includegraphics[width=0.96\linewidth]{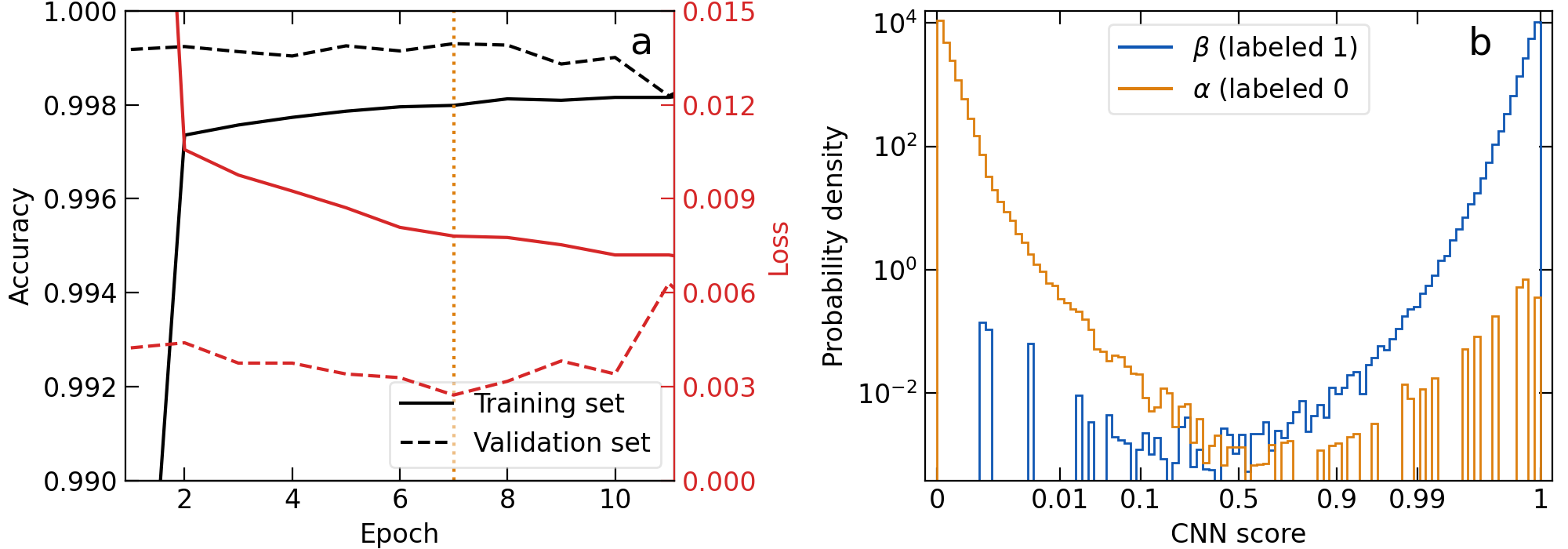}
\caption{(a) Accuracy (black) or loss (red) during the epoch. Solid and dot lines represent the training set and validation set, respectively. CNN model is selected at epoch 7 (orange dashed line) to prevent overfitting. (b) Score distribution in the validation set. Bins are spaced using the error function: $0.5\cdot[1+\mathrm{erf}\left((x-0.5)/0.18\right)]$ for equally spaced $x$ between 0 and 1, to effectively show the dense distribution at each end.}
\label{fig:training_result}
\end{figure}

Overfitting may occur when the loss of the training set decreases while the loss of the validation grows. It is also considered overfitting and the training is stopped when the validation loss does not decrease within five epochs.
The optimized weights are chosen at the epoch right before the overfitting happens. We expect the overfitting to occur by the fortieth epoch and set the hard limit there. As shown in figure~\ref{fig:training_result}, the training does not improve the validation loss after the seventh epoch. As a result, we obtained 99.93\% accuracy for the validation data set at epoch 7.

%% file: 4.results.tex
\section{Results}
\label{sec:result}

\subsection{Score cut and efficiency}
With the conventional PSD method, our selection strategy for the IBD candidate events is to maintain the selection efficiency for the electron recoil events at a value higher than 99\% and to have moderate rejection efficiencies for the fast-$n$ background throughout the energy range of interest (1-10 MeV).
It is to minimize any distortion and systematic uncertainty in the IBD prompt energy spectrum caused by the selection efficiency.
This strategy, however, may introduce an additional systematic error for the unknown difference between the measured background from reactor-off periods and the background mixed with IBD in the reactor-on data. 
Remained fast-$n$ background rate by an inefficient rejection, especially at low energy, can change in time due to an unstable detector response~\cite{jinyu_kim_2022_6680618} or by fluctuations of the environmental condition~\cite{STEREO:2019ztb}.
In the NEOS-I case with the conventional PSD method, a 10\% conservative error for the reactor-off spectrum was assigned due to this unknown background fluctuation~\cite{NEOS:2016wee}.

\begin{figure*}[htb!]%
\centering
\includegraphics[width=0.99\textwidth]{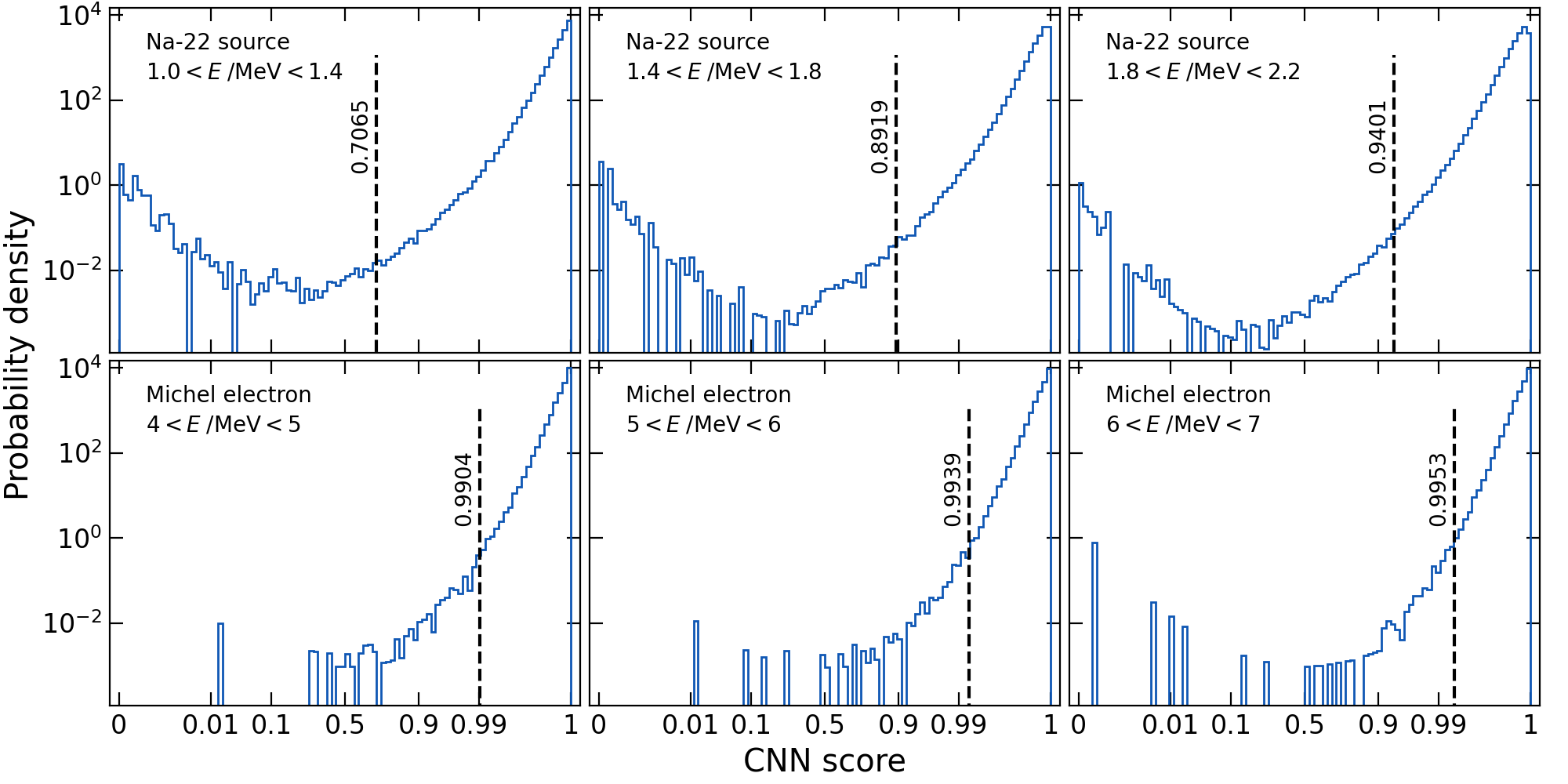}
\caption{CNN score distributions of $^{22}$Na source data (top) and Michel electron events (bottom) at different energy ranges. Binning is the same as in figure~\ref{fig:training_result}b. The black dashed lines and the numbers indicate the score cuts satisfying 99.5\% $\gamma$ selection efficiency.
}%
\label{fig:Scorecut} %
\end{figure*}

Our goal in employing CNN for IBD selection is to maintain the same high selection efficiency for the electron recoil events while improving the fast-$n$ rejection efficiency.
The selection efficiency is investigated using evaluated CNN scores of the clean electron recoil events from the $^{22}$Na source calibration data and the Michel electron data for the low and high energy ranges, respectively.
In this work, the data are sorted into different energy windows and a score of 99.5\% acceptance for each energy window is found, as shown in figure~\ref{fig:Scorecut}.
A score cut as a function of energy is found by interpolation of those score points. The cut value gets close to 1 as the energy increases.

As mentioned in section~\ref{ssec:data_samples}, clean fast-$n$-only events, especially at low energy, cannot be sorted out in the NEOS data.
Estimation of efficiency with the CNN score is not as easy for the fast-$n$ rejection as for the $\beta$/$\gamma$ selection.
Therefore, the CNN score cut in this work is validated by checking consistency among a result by CNN, another by conventional PSD, and the other without PSD.
Since the $\beta$/$\gamma$ selection efficiencies were set to be higher than 99\% in both the CNN and the conventional methods, the reactor on-off spectra from all three methods should agree.

\subsection{Application to the IBD selection}
\begin{figure}[!htb]
\centering
\includegraphics[width=\textwidth]{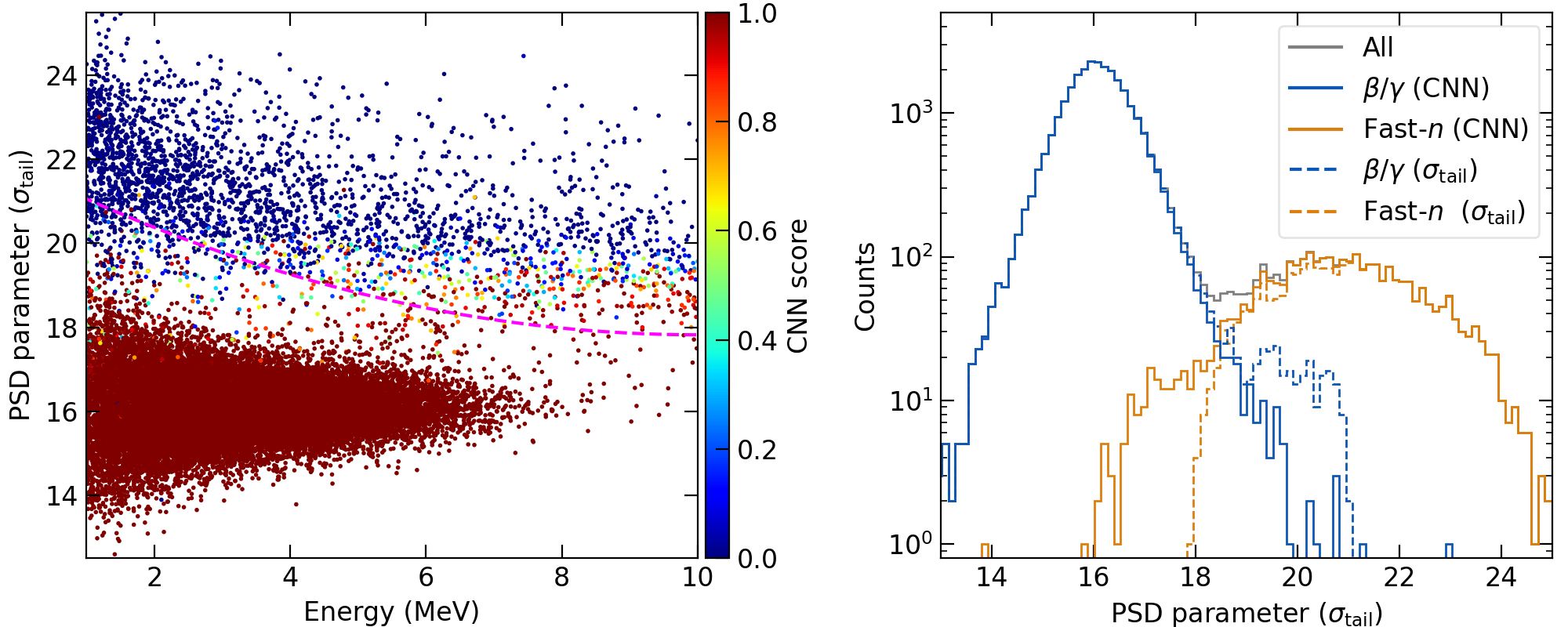}
\caption{The left figure shows the PSD parameter ($\sigma_{\mathrm{tail}}$) versus the energy of the IBD prompt event candidates in the early reactor-on data of NEOS-II.
Color represents the evaluated CNN score. The magenta dashed curve indicates a PSD cut on the conventional PSD parameter ($\sigma_{\mathrm{tail}}$).
The right figure shows the projected data on $\sigma_{\mathrm{tail}}$.
Blue and orange histograms of solid (dashed) lines are selected $\beta$/$\gamma$ and fast-$n$ events, respectively, by the CNN score ($\sigma_{\mathrm{tail}}$) cut.}
\label{fig:ibd_score}
\end{figure}

PSD parameter~($\sigma_{\mathrm{tail}}$) of the IBD prompt event candidates from an early reactor-on data set is shown in figure~\ref{fig:ibd_score}
The $\sigma_{\mathrm{tail}}$ of $\beta$/$\gamma$ events lies along around 16, and the resolution becomes poorer as energy decreases.
The $\sigma_{\mathrm{tail}}$ cut value along with the fast-$n$'s $\sigma_{\mathrm{tail}}$ decreases as energy grows. The cut for this early data set seems quite loose. Still, it is to sustain the same efficiency with the following data sets with poorer discrimination caused by a light yield degradation during the NEOS-II measurement~\cite{jinyu_kim_2022_6680618}.
In terms of CNN score, most events are separated into scores very close to 0 or 1, and not many events exist in the middle.
Some events on the high-energy side and above the $\sigma_{\mathrm{tail}}$ cut get scores close to half, mainly because the CNN is optimized using only the low-energy Bi-Po events.
However, events with dark red color can only survive after the CNN cut for the high-energy side because the cut is at a score higher than 0.99 for $E>4$~MeV, as shown in figure~\ref{fig:Scorecut}.
On the low energy side, in contrast, a considerable number of events below the $\sigma_{\mathrm{tail}}$ cut are classified clearly as fast-$n$ events.
In short, the CNN score cut rejects more events, many of which are in the low-energy region.

\begin{figure}[htb!]\textbf{}
\centering
\includegraphics[width=0.9\textwidth]{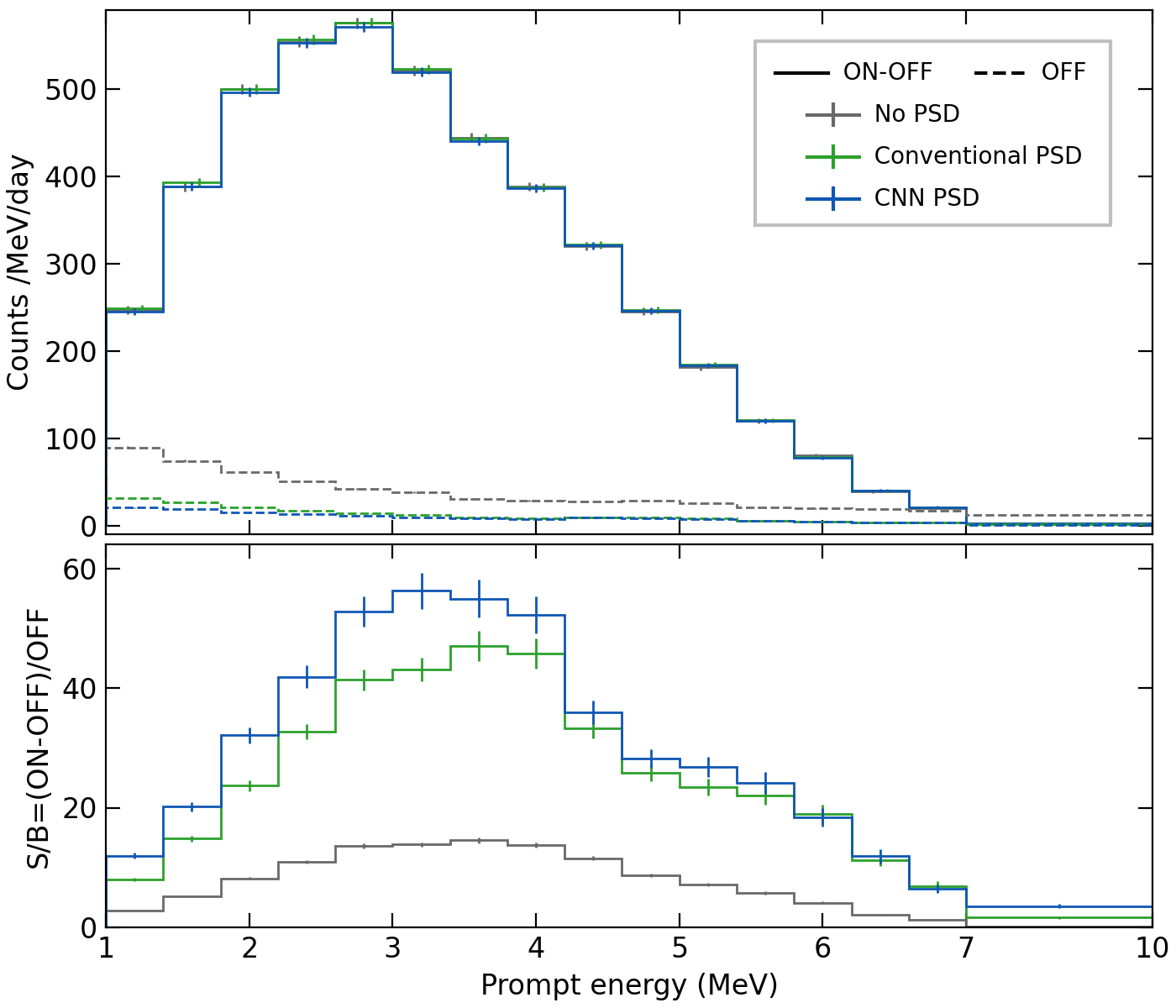}
\caption{Top: IBD signal (ON-OFF; solid line) and background (OFF; dashed line) spectra with different analysis methods--no PSD (gray), conventional PSD (green), and PSD using CNN (blue).
Bottom: Signal-to-background ratio for each analysis method.
}
\label{fig:finalresult}
\end{figure}

Figure~\ref{fig:finalresult} shows the IBD prompt energy distributions of the signals (reactor-on minus reactor-off) and backgrounds (reactor-off) with three analysis methods: one without PSD, another with $\sigma_{\mathrm{tail}}$ cut, and the other by the CNN score cut.
Both PSD methods show significant backgroud reductions, while the CNN method rejects more background at low energy.
Despite significant differences among the background spectra, subtracting them from their corresponding reactor-on spectra leaves the IBD signal spectra with almost perfect agreements among one another. The total signal rate (on$-$off) after the $\sigma_{\mathrm{tail}}$ cut is 100\%, and the one by CNN is 99.8\%, compared to the result of 1864 events per day without PSD, respectively.
Daily background rates from CNN and $\sigma_{\mathrm{tail}}$ methods are 59.9$\pm$0.7 and 76.2$\pm$0.8, respectively, enhancing the signal-to-background ratio by more than 20\%, from 24 to 31.
As intended and expected, improvement is more significant on the low-energy side as CNN is trained with low-energy data.

%% file: 5.summary.tex
\section{Summary and discussion}
\label{sec:summary}

This study shows that CNN helps PSD to improve fast-$n$ background rejection by a considerable amount, at least in the case that a suitable data set exists for training in the region where the conventional PSD shows limited power.
Excellent separation between $\beta$/$\gamma$ and fast-$n$/$\alpha$ events may let us avoid complicated cut design and efficiency estimation in the conventional method.
At the same time, it takes a similar time and effort to optimize the CNN architecture.
It is also shown that a normalized raw FFT power spectrum can be a good input for CNN without much loss of key features of recoil event type and  avoiding bias.
\begin{figure}[!htb]
\centering
\includegraphics[width=0.9\textwidth]{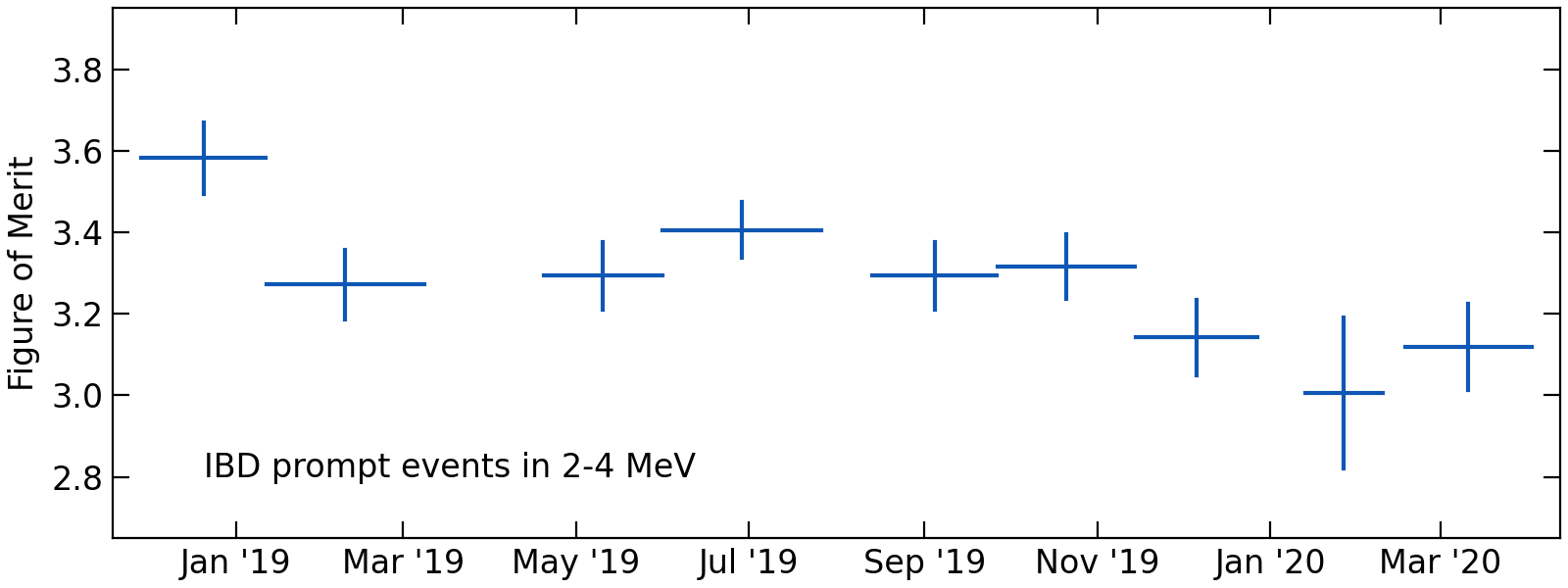}
\caption{The figure of merit (FoM) for the PSD parameter for the prompt events of 2-4 MeV energy during the reactor-on period of the NEOS-II data taking. The decrease in FoM is due to the continuous light yield degradation. The data are divided into 9 groups and the first data group of 44 days and preceding reactor-off data for 29 days are shown in this study.}\label{fig:fom_ibd}
\end{figure}

The result presented here is for the first 29+44 days with the reactor off+on, respectively, out of the total 500 days of NEOS-II data taking.
The FoM of the PSD parameter became lower in time as shown in figure~\ref{fig:fom_ibd}, because of the degradation of the scintillation light yield during the whole NEOS-II data-taking period, as reported in \cite{jinyu_kim_2022_6680618}.
Therefore, data are divided into 9 groups for the reactor-on period, and each data group is trained and evaluated separately using the same process developed in this study. 
A fine-tuning of the CNN score cut for each data group is going on to keep the selection efficiency to be as stable and close to unity as possible to minimize the uncertainty from the efficiency fluctuation.
A preliminary result for the whole NEOS-II data set has been presented in ~\cite{jinyu_kim_2022_6680618}.
One of the primary goals of NEOS-II is to differentiate the IBD spectrum of $^{235}$U and that of plutonium.
This spectrum decomposition is derived from the change of the total spectrum in time, which is only a few percent for a reactor operation cycle.
Therefore, small improvements in statistical and systematic accuracy will help improve the experimental result more than a little.
However, the signal-to-background ratio is already as large as 24 without the help of CNN.

A limited feature of this study is that only low visible energy $\alpha$ events are used in the CNN training to mitigate the fast-$n$ background whose visible energy range spans all the IBD prompt energy ranges and above.
Further PSD improvement is expected to be achieved, at least by getting clean fast-$n$ and $\beta$/$\gamma$ data samples at the rest of the energy range.

%% file: main.bbl
\providecommand{\href}[2]{#2}\begingroup\raggedright\begin{thebibliography}{10}

\bibitem{Vogel:1999zy}
P.~Vogel and J.F.~Beacom, \emph{{Angular distribution of neutron inverse beta
  decay, anti-neutrino(e) + p ---\ensuremath{>} e+ + n}},
  \href{https://doi.org/10.1103/PhysRevD.60.053003}{\emph{Phys. Rev. D}
  {\bfseries 60} (1999) 053003}
  [\href{https://arxiv.org/abs/hep-ph/9903554}{{\ttfamily hep-ph/9903554}}].

\bibitem{Mention_2011}
G.~Mention, M.~Fechner, T.~Lasserre, T.A.~Mueller, D.~Lhuillier, M.~Cribier
  et~al., \emph{Reactor antineutrino anomaly},
  \href{https://doi.org/10.1103/physrevd.83.073006}{\emph{Physical Review D}
  {\bfseries 83} (2011) }.

\bibitem{Davis:2016gkz}
J.R.~Davis, E.~Brubaker and K.~Vetter, \emph{{Fast neutron background
  characterization with the Radiological Multi-sensor Analysis Platform
  (RadMAP)}}, \href{https://doi.org/10.1016/j.nima.2017.03.042}{\emph{Nucl.
  Instrum. Meth. A} {\bfseries 858} (2017) 106}
  [\href{https://arxiv.org/abs/1611.04996}{{\ttfamily 1611.04996}}].

\bibitem{STEREO:2019ztb}
{\scshape STEREO} collaboration, \emph{{Improved sterile neutrino constraints
  from the STEREO experiment with 179 days of reactor-on data}},
  \href{https://doi.org/10.1103/PhysRevD.102.052002}{\emph{Phys. Rev. D}
  {\bfseries 102} (2020) 052002}
  [\href{https://arxiv.org/abs/1912.06582}{{\ttfamily 1912.06582}}].

\bibitem{5485131}
G.~Liu, M.J.~Joyce, X.~Ma and M.D.~Aspinall, \emph{A digital method for the
  discrimination of neutrons and $\gamma$ rays with organic scintillation
  detectors using frequency gradient analysis},
  \href{https://doi.org/10.1109/TNS.2010.2044246}{\emph{IEEE Transactions on
  Nuclear Science} {\bfseries 57} (2010) 1682}.

\bibitem{BALMER2015146}
M.J.~Balmer, K.A.~Gamage and G.C.~Taylor, \emph{Comparative analysis of pulse
  shape discrimination methods in a 6li loaded plastic scintillator},
  \href{https://doi.org/https://doi.org/10.1016/j.nima.2015.03.089}{\emph{Nuclear
  Instruments and Methods in Physics Research Section A: Accelerators,
  Spectrometers, Detectors and Associated Equipment} {\bfseries 788} (2015)
  146}.

\bibitem{HUBBARD201964}
M.~Hubbard, M.~Taggart and P.~Sellin, \emph{Exploration of Fourier based
  algorithms and detector designs for pulse shape discrimination},
  \href{https://doi.org/https://doi.org/10.1016/j.nima.2019.03.020}{\emph{Nuclear
  Instruments and Methods in Physics Research Section A: Accelerators,
  Spectrometers, Detectors and Associated Equipment} {\bfseries 930} (2019)
  64}.

\bibitem{Lenet5}
Y.~Lecun, L.~Bottou, Y.~Bengio and P.~Haffner, \emph{Gradient-based learning
  applied to document recognition},
  \href{https://doi.org/10.1109/5.726791}{\emph{Proceedings of the IEEE}
  {\bfseries 86} (1998) 2278}.

\bibitem{NIPS2012_c399862d}
A.~Krizhevsky, I.~Sutskever and G.E.~Hinton, \emph{Imagenet classification with
  deep convolutional neural networks}, {\emph{Advances in neural information
  processing systems} {\bfseries 25} (2012)}.

\bibitem{https://doi.org/10.48550/arxiv.1409.1556}
K.~Simonyan and A.~Zisserman, \emph{Very deep convolutional networks for
  large-scale image recognition},  2014.
\newblock 10.48550/arxiv.1409.1556.

\bibitem{https://doi.org/10.48550/arxiv.1409.4842}
C.~Szegedy, W.~Liu, Y.~Jia, P.~Sermanet, S.~Reed, D.~Anguelov et~al.,
  \emph{Going deeper with convolutions},  2014.
\newblock 10.48550/ARXIV.1409.4842.

\bibitem{https://doi.org/10.48550/arxiv.1512.03385}
K.~He, X.~Zhang, S.~Ren and J.~Sun, \emph{Deep residual learning for image
  recognition},  2015.
\newblock 10.48550/ARXIV.1512.03385.

\bibitem{Aurisano:2016jvx}
A.~Aurisano, A.~Radovic, D.~Rocco, A.~Himmel, M.D.~Messier, E.~Niner et~al.,
  \emph{{A Convolutional Neural Network Neutrino Event Classifier}},
  \href{https://doi.org/10.1088/1748-0221/11/09/P09001}{\emph{JINST} {\bfseries
  11} (2016) P09001} [\href{https://arxiv.org/abs/1604.01444}{{\ttfamily
  1604.01444}}].

\bibitem{DUNE:2020gpm}
{\scshape DUNE} collaboration, \emph{{Neutrino interaction classification with
  a convolutional neural network in the DUNE far detector}},
  \href{https://doi.org/10.1103/PhysRevD.102.092003}{\emph{Phys. Rev. D}
  {\bfseries 102} (2020) 092003}
  [\href{https://arxiv.org/abs/2006.15052}{{\ttfamily 2006.15052}}].

\bibitem{MicroBooNE:2016dpb}
{\scshape MicroBooNE} collaboration, \emph{{Convolutional Neural Networks
  Applied to Neutrino Events in a Liquid Argon Time Projection Chamber}},
  \href{https://doi.org/10.1088/1748-0221/12/03/P03011}{\emph{JINST} {\bfseries
  12} (2017) P03011} [\href{https://arxiv.org/abs/1611.05531}{{\ttfamily
  1611.05531}}].

\bibitem{MicroBooNE:2018kka}
{\scshape MicroBooNE} collaboration, \emph{{Deep neural network for pixel-level
  electromagnetic particle identification in the MicroBooNE liquid argon time
  projection chamber}},
  \href{https://doi.org/10.1103/PhysRevD.99.092001}{\emph{Phys. Rev. D}
  {\bfseries 99} (2019) 092001}
  [\href{https://arxiv.org/abs/1808.07269}{{\ttfamily 1808.07269}}].

\bibitem{MicroBooNE:2020yze}
{\scshape MicroBooNE} collaboration, \emph{{Semantic segmentation with a sparse
  convolutional neural network for event reconstruction in MicroBooNE}},
  \href{https://doi.org/10.1103/PhysRevD.103.052012}{\emph{Phys. Rev. D}
  {\bfseries 103} (2021) 052012}
  [\href{https://arxiv.org/abs/2012.08513}{{\ttfamily 2012.08513}}].

\bibitem{NEXT:2016ire}
{\scshape NEXT} collaboration, \emph{{Background rejection in NEXT using deep
  neural networks}},
  \href{https://doi.org/10.1088/1748-0221/12/01/T01004}{\emph{JINST} {\bfseries
  12} (2017) T01004} [\href{https://arxiv.org/abs/1609.06202}{{\ttfamily
  1609.06202}}].

\bibitem{DBLP:journals/corr/abs-1809-06166}
N.~Choma, F.~Monti, L.~Gerhardt, T.~Palczewski, Z.~Ronaghi, Prabhat et~al.,
  \emph{Graph neural networks for IceCube signal classification}, {\emph{CoRR}
  {\bfseries abs/1809.06166} (2018) }
  [\href{https://arxiv.org/abs/1809.06166}{{\ttfamily 1809.06166}}].

\bibitem{7838264}
E.~Racah, S.~Ko, P.~Sadowski, W.~Bhimji, C.~Tull, S.-Y.~Oh et~al.,
  \emph{Revealing fundamental physics from the Daya Bay neutrino experiment
  using deep neural networks},  in \emph{2016 15th IEEE International
  Conference on Machine Learning and Applications (ICMLA)}, pp.~892--897, 2016,
  \href{https://doi.org/10.1109/ICMLA.2016.0160}{DOI}.

\bibitem{Domine:2019zhm}
{\scshape DeepLearnPhysics} collaboration, \emph{{Scalable deep convolutional
  neural networks for sparse, locally dense liquid argon time projection
  chamber data}},
  \href{https://doi.org/10.1103/PhysRevD.102.012005}{\emph{Phys. Rev. D}
  {\bfseries 102} (2020) 012005}
  [\href{https://arxiv.org/abs/1903.05663}{{\ttfamily 1903.05663}}].

\bibitem{KamLAND-Zen:2016pfg}
{\scshape KamLAND-Zen} collaboration, \emph{{Search for Majorana Neutrinos near
  the Inverted Mass Hierarchy Region with KamLAND-Zen}},
  \href{https://doi.org/10.1103/PhysRevLett.117.082503}{\emph{Phys. Rev. Lett.}
  {\bfseries 117} (2016) 082503}
  [\href{https://arxiv.org/abs/1605.02889}{{\ttfamily 1605.02889}}].

\bibitem{Griffiths:2018zde}
J.~Griffiths, S.~Kleinegesse, D.~Saunders, R.~Taylor and A.~Vacheret,
  \emph{{Pulse Shape Discrimination and Exploration of Scintillation Signals
  Using Convolutional Neural Networks}},
  \href{https://arxiv.org/abs/1807.06853}{{\ttfamily 1807.06853}}.

\bibitem{MINERvA:2018smv}
{\scshape MINERvA} collaboration, \emph{{Reducing model bias in a deep learning
  classifier using domain adversarial neural networks in the MINERvA
  experiment}},
  \href{https://doi.org/10.1088/1748-0221/13/11/P11020}{\emph{JINST} {\bfseries
  13} (2018) P11020} [\href{https://arxiv.org/abs/1808.08332}{{\ttfamily
  1808.08332}}].

\bibitem{NEOS:2016wee}
{\scshape NEOS} collaboration, \emph{{Sterile Neutrino Search at the NEOS
  Experiment}},
  \href{https://doi.org/10.1103/PhysRevLett.118.121802}{\emph{Phys. Rev. Lett.}
  {\bfseries 118} (2017) 121802}
  [\href{https://arxiv.org/abs/1610.05134}{{\ttfamily 1610.05134}}].

\bibitem{Kim_2015}
B.R.~Kim, B.Y.~Han, E.J.~Jeon, K.K.~Joo, J.~Kang, N.~Khan et~al., \emph{Pulse
  shape discrimination capability of metal-loaded organic liquid scintillators
  for a short-baseline reactor neutrino experiment},
  \href{https://doi.org/10.1088/0031-8949/90/5/055302}{\emph{Physica Scripta}
  {\bfseries 90} (2015) 055302}.

\bibitem{NEOS:2015dzs}
{\scshape NEOS} collaboration, \emph{{Development and Mass Production of a
  Mixture of LAB- and DIN-based Gadolinium-loaded Liquid Scintillator for the
  NEOS Short-baseline Neutrino Experiment}},
  \href{https://doi.org/10.1007/s10967-016-4826-1}{\emph{J. Radioanal. Nucl.
  Chem.} {\bfseries 310} (2016) 311}
  [\href{https://arxiv.org/abs/1511.05551}{{\ttfamily 1511.05551}}].

\bibitem{yoomin:2018nu}
{\scshape NEOS} collaboration, Y.M.~Oh et~al., \emph{NEOS result and
  prospects},  June 2018.
\newblock 10.5281/zenodo.1286994.

\bibitem{NEOSII:2020gai}
{\scshape NEOS II} collaboration, \emph{{Pulse-shape Discrimination of Fast
  Neutron Background using Convolutional Neural Network for NEOS II}},
  \href{https://doi.org/10.3938/jkps.77.1118}{\emph{J. Korean Phys. Soc.}
  {\bfseries 77} (2020) 1118}
  [\href{https://arxiv.org/abs/2009.13355}{{\ttfamily 2009.13355}}].

\bibitem{Michel:1949qe}
L.~Michel, \emph{{Interaction between four half spin particles and the decay of
  the $\mu$ meson}},
  \href{https://doi.org/10.1088/0370-1298/63/5/311}{\emph{Proc. Phys. Soc. A}
  {\bfseries 63} (1950) 514}.

\bibitem{Goodfellow-et-al-2016}
I.~Goodfellow, Y.~Bengio and A.~Courville, \emph{Deep Learning}, MIT Press
  (2016).

\bibitem{Asurvey}
A.~Khan, A.~Sohail, U.~Zahoora and A.~Saeed, \emph{A survey of the recent
  architectures of deep convolutional neural networks},
  \href{https://doi.org/10.1007/s10462-020-09825-6}{\emph{Artificial
  Intelligence Review} {\bfseries 53} (2020)}.

\bibitem{rala2021neural}
J.~Rala~Cordeiro, A.~Raimundo, O.~Postolache and P.~Sebasti{\~a}o, \emph{Neural
  architecture search for 1d cnns—different approaches tests and
  measurements}, \href{https://doi.org/10.3390/s21237990}{\emph{Sensors}
  {\bfseries 21} (2021) 7990}.

\bibitem{chollet2015keras}
F.~Chollet et~al., \emph{Keras},  2015.

\bibitem{DBLP:journals/corr/AbadiBCCDDDGIIK16}
M.~Abadi, P.~Barham, J.~Chen, Z.~Chen, A.~Davis, J.~Dean et~al.,
  \emph{Tensorflow: {A} system for large-scale machine learning}, {\emph{CoRR}
  {\bfseries abs/1605.08695} (2016) }
  [\href{https://arxiv.org/abs/1605.08695}{{\ttfamily 1605.08695}}].

\bibitem{https://doi.org/10.48550/arxiv.1412.6980}
D.P.~Kingma and J.~Ba, \emph{Adam: A method for stochastic optimization},
  2014.
\newblock 10.48550/ARXIV.1412.6980.

\bibitem{jinyu_kim_2022_6680618}
J.~Kim, \emph{Sterile neutrino i\_neos-ii new results},  June 2022.
\newblock 10.5281/zenodo.6680618.

\end{thebibliography}\endgroup
